\documentclass[prl,twocolumn,amsfonts,showpacs]{revtex4}
\usepackage{graphics,bm,natbib}
\begin{document}
\widetext
\title{Charge Ordering Due to Magnetic Symmetry Breaking}
\author{I. V. Solovyev}
\altaffiliation[Present address: ] {Institute for Solid State Physics,
University of Tokyo, Kashiwanoha 5-1-5, Kashiwa 277-8531, Japan}
\email[Electronic address: ]{igor@issp.u-tokyo.ac.jp}
\affiliation{ Tokura Spin Superstructure Project,
ERATO Japan Science and Technology Corporation, \\
c/o National Institute of Advanced Industrial Science and Technology, \\
Central 4, 1-1-1 Higashi, Tsukuba, Ibaraki 305-8562, Japan
}
\date{\today}

\widetext
\begin{abstract}
It is argued that both transitions observed in 50\%-doped manganites, at the
N\'{e}el temperature ($T_{\rm N}$) and the so-called charge ordering temperature
($T_{\rm CO}$), are magnetic. $T_{\rm N}$ corresponds to
the order-disorder transition, which takes place between ferromagnetic
zigzag chains, while
the coherent motion of spins within the chains is
destroyed only around $T_{\rm CO}$.
The magnetic structure realized below $T_{\rm CO}$ is highly anisotropic.
It is dressed by the lattice distortion and leads to
huge anisotropy of the electronic structure, which explains stability of
this state as well as the form of
charge-orbital pattern above $T_{\rm N}$.
The type of phase transition at $T_{\rm N}$
depends on lattice interactions.
\end{abstract}

\pacs{75.25+z, 75.30.Vn, 71.23.-k, 75.10.Lp}


\maketitle

\begin{figure}[h!]
\centering \noindent
\resizebox{5.0cm}{!}{\includegraphics{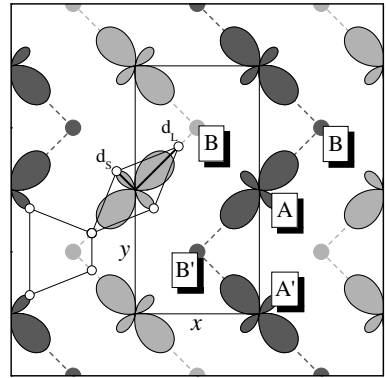}}
\caption{Magnetic structures in the CO phase.
Two AFM sublattices below $T_{\rm N}$ are shown by different colors. Above $T_{\rm N}$,
the N\'{e}el state is replaced by the inter-chain disorder, which coexists with
nearly perfect FM order within the chains. New magnetic unit cell
is denoted by $x$-$y$. The anisotropic magnetic structures give rise to the OO patter,
which is coupled with the oxygen displacements, shown by open
circles. The CO is only illusive and reflected in the existence of two
non-equivalent sites ($A$ and $B$), produced by the magnetic symmetry breaking.}
\label{fig.pattern}
\end{figure}

  The so-called ''charge-ordered'' state of the colossal magnetoresistive perovskite
manganites is one of the most interesting, important, and at the same
time -- somewhat obscured phenomena \cite{Tokura}. Since the charge ordering (CO)
is accompanied by the orbital ordering (OO), and sometimes -- by the very
peculiar zigzag antiferromagnetic (AFM) ordering (Fig.~\ref{fig.pattern}),
it always causes a problem with the identification of
the true order parameter
responsible for the physics of this complicated phase.
Despite the long
history, a visible progress in
this direction
has been made only recently \cite{Hotta,PRL99,PRB01,Brink,Yunoki}.
It appeared that many low-temperature ($T$) properties of CO
manganites, like Nd$_{1-x}$Ca$_x$MnO$_3$ (NCMO),
Nd$_{1-x}$Sr$_x$MnO$_3$ (NSMO),
and La$_{1-x}$Sr$_{1+x}$MnO$_4$ (LSMO) at $x$$=$$0.5$, can be understood
on the basis of purely magnetic considerations,
by combining traditional ideas of the double exchange (DE) \cite{deGennes}
with the anisotropy of transfer interactions between two $e_g$ orbitals \cite{Brink2},
and formally without
involving additional mechanisms of the CO and OO.

  It is true that in order to stabilize the zigzag AFM state,
one should first break the cubic symmetry. Most
of attempts were devoted to the search of
some \textit{external}
(with respect to the DE) mechanism, which would first stabilize
the CO/OO pattern shown in Fig.~\ref{fig.pattern}, and thereby predetermine
the form of the zigzag AFM state.
The main idea proposed in Refs.~\cite{PRL99,PRB01} was that
in the degenerate DE model, the cubic symmetry can be broken by \textit{internal}
means, by imposing the zigzag AFM order which forms the desired CO/OO ordering
pattern, and this symmetry breaking is \textit{already sufficient} to stabilize the
zigzag AFM state.
Thus, the appearance of the CO phase  at $T$$=$$0$ can be explained
by the following formula: anisotropic AFM order $\rightarrow$
symmetry breaking $\rightarrow$ stability of this AFM order. The last step is
considerably facilitated by the fact that the zigzag AFM phase is a band
insulator~\cite{Hotta,PRL99,PRB01,Brink}. The finding is very important,
because the magnetic origin of the CO phase naturally explains
the fact that it can be melted by the
magnetic field, giving rise to negative
colossal magnetoresistance (CMR) effect~\cite{melting}.
The second mechanism, which
additionally stabilizes the zigzag AFM state~\cite{PRB01,Yunoki}, is the
Jahn-Teller (JT) effect~\cite{Millis}.

  Is this the end of the story? Apparently not,
because this
picture is believed to be at odds
with the temperature behavior of CO manganites, which can be of
two types. The first one, realized in NCMO and
LSMO, is characterized by
the distinct transition
temperature, $T_{\rm CO}$, which is manifested in appearance of the
CO/OO
superlattice spots in the neutron \cite{Sternlieb} and x-ray \cite{Murakami}
experiments, as well as
the sharp increase of the resistivity, and the peak of magnetic
susceptibility \cite{Tokura,Tomioka}. The
N\'{e}el temperature, $T_{\rm N}$, is mainly manifested in the appearance of
AFM superlattice spots \cite{Sternlieb,Murakami}.
The typical values of $T_{\rm CO}$ and $T_{\rm N}$, observed in
LSMO, are 217 and 110 K, respectively.
Since $T_{\rm CO}$$>$$T_{\rm N}$,
the CO is typically regarded as a more fundamental effect,
which drives the AFM transition at $T_{\rm N}$.
Another possibility, realized in NSMO, is the first-order transition to the
metallic ferromagnetic
(FM) phase
above
$T_{\rm N}$, which coincides with $T_{\rm CO}$. More generally,
$x$$=$$0.5$ is the point of multi-critical behavior of
CMR manganites~\cite{Tomioka}.

  The purpose of this Letter is to argue that such a temperature behavior can be
understood in terms of the partial spin disorder. Namely,
only the long-range AFM order in the $x$-direction disappears at $T_{\rm N}$, whereas
the FM coupling within the chains survives this transition.
Thus, in the interval
$T_{\rm N}$$<$$T$$<$$T_{\rm CO}$
we deal with the one-dimensional spin disorder in the
direction perpendicular to the chains.
The disorder is highly anisotropic.
It breaks the cubic symmetry, stabilizes the FM chains and gives rise to the
CO/OO pattern even above $T_{\rm N}$ \cite{Sternlieb,Murakami}.
We will concentrate on the dynamics of two-dimensional pattern
shown in Fig.~\ref{fig.pattern}.

  The behavior of CO manganites at $T$$=$$0$
is well described by the following
model, combining DE physics with the harmonic JT effect~\cite{Yunoki}:
\begin{equation}
\widehat{\cal H}_{\bf ij} = - \xi_{\bf ij} \widehat{t}_{\bf ij} +
\left( \widehat{\Delta}_{\bf i} \delta_{\rm JT}
+ \frac{\kappa}{2} \delta^2_{\rm JT}
- J^S \zeta_{\bf i} \right) \delta_{\bf ij},
\label{eqn:H}
\end{equation}
where $\xi_{\bf ij}$$=$$ \left( \frac{1+{\bf e_i} \cdot {\bf e_j}}{2} \right)^{1/2}$
\cite{comment.2}.
${\bf e_i}$ is the spin direction at the site ${\bf i}$.
$\widehat{t}_{\bf ij}$$=$$\| t_{\bf ij}^{LL'} \|$
is the matrix of transfer interactions between nearest-neighbor (NN) $e_g$ orbitals:
$x^2$-$y^2$ ($L,L'$$=$$1$) and $3z^2$-$r^2$ ($L,L'$$=$$2$), in the local
frame specified by the oxygen sites (corresponding to the
$xy$ and $3z^2$-$r^2$ orbitals in the $xy$-frame in Fig.~\ref{fig.pattern}).
The NN hoppings along the $(\pm1,1)$ directions are parameterized according
to Slater and Koster:
$\widehat{t}_{\bf ij}$$=$$\frac{t}{4}(2$$\mp$$\sqrt{3}\widehat{\sigma}_z$$
+\widehat{\sigma}_z)$, where $\widehat{\sigma}_x$ and $\widehat{\sigma}_z$
are the Pauli matrices.
$\widehat{\Delta}_{\bf i}$$=$$\| \Delta_{\bf i}^{LL'} \|$ describes
the $e_g$ levels splitting caused by the cooperative JT distortion
$\delta_{\rm JT}$$=$$(d_L$$-$$d_S)/(d_L$$+$$d_S)$:
$\widehat{\Delta}_{A}$$=$$-$$\frac{\lambda}{2}(\sqrt{3}\widehat{\sigma}_x$$
+$$\widehat{\sigma}_z)$,
$\widehat{\Delta}_{A'}$$=$$\frac{\lambda}{2}(\sqrt{3}\widehat{\sigma}_x$$
-$$\widehat{\sigma}_z)$, and
$\widehat{\Delta}_{B}$$=$$-$$\frac{\lambda}{2}\widehat{\sigma}_z$
($\lambda$ being the electron-phonon coupling constant).
The considered distortion is compatible with the magnetic zigzag structure
(Fig.~\ref{fig.pattern}).
$t(\delta_{\rm JT})$
and $\lambda$ are estimated
from Harrison's law for the $p$-$d$ transfer interactions~\cite{PRB01}.
This yields $\lambda$$\simeq$$14 t_0$.
$J^S$$<$$0$ is the AFM superexchange (SE) coupling between NN spins
($|J^S|$$\sim$$0.1 t_0$~\cite{Springer}) and
$2\zeta_{\bf i}$$=$$\sum_{{\bf n} \in {\rm NN}} {\bf e_i} \cdot {\bf e_n}$.
The spring constant $\kappa$ may vary from several tens to
hundreds eV~\cite{Millis} (below we will give a more realistic estimate
for $\kappa$ from the analysis of the phase transition at
$T_{\rm N}$).
$t_0$$=$$t(0)$
is used as the energy unit
($t_0$$=$$0.7$ eV \cite{Springer}).

  What goes first in the
Hamiltonian (\ref{eqn:H}) and causes the required anisotropy of the
electronic structure: the magnetism or the JT effect?
In the harmonic approximation around
a homogeneous magnetic state, the ground state may have only one minimum
at $\delta_{\rm JT}$$=$$0$~\cite{LAZ}.
Therefore, the JT distortion alone cannot be the source of the
anisotropy. On the other hand, the magnetic part of the Hamiltonian (\ref{eqn:H})
can, and the equilibrium magnetic structure at $T$$=$$0$ can be
successfully described in terms of competition between DE and SE
interactions~\cite{Hotta,PRL99,PRB01,Brink}.
Therefore, even for the finite-temperature description, it seems to be
logical to start with purely magnetic Hamiltonian by \textit{enforcing}
$\delta_{\rm JT}$$=$$0$ in Eq.~(\ref{eqn:H}). What do we expect
in this case?

  The stability condition of the zigzag AFM phase at $T$$=$$0$
is given by
$J_{AB}^{\uparrow \downarrow}$$<$$|J^S|$$<$$J_{AB}^{\uparrow \uparrow}$,
where $J_{AB}^{\uparrow \uparrow}$ ($J_{AB}^{\uparrow \downarrow}$) are the
NN DE interactions within the same (between different) zigzag chains~\cite{PRB01}.
For $\delta_{\rm JT}$$=$$0$
the exchange anisotropy
$\Delta J$$=$$J_{AB}^{\uparrow \uparrow}$$-$$J_{AB}^{\uparrow \downarrow}$
is of the order of $0.07$ (and becomes larger for $\delta_{\rm JT}$$>$$0$).
Our basic idea is to connect this anisotropy,
which roughly corresponds to
the temperature range of
550 K,
with the existence of two transition points, $T_{\rm N}$ and $T_{\rm CO}$.
Since $|J^S|$ is close to $J_{AB}^{\uparrow \downarrow}$
(about $0.106$ for $\delta_{\rm JT}$$=$$0$), the AFM coupling between the chains,
$J_{AB}^{\uparrow \downarrow}$$+$$J^S$, will be much weaker than the
FM interaction within the chain, $J_{AB}^{\uparrow \uparrow}$$+$$J^S$.
Therefore,
the long-range AFM
order in the $x$-direction will be destroyed first with the increase of $T$.
This corresponds to the first transition point, $T_{\rm N}$, and
yields the following model for the spin disorder below $T_{\rm N}$.

  All spins in an arbitrary chain $\ell$ are rigidly coupled by
$J_{AB}^{\uparrow \uparrow}$ (the effect of inter-chain disorder
on $J_{AB}^{\uparrow \uparrow}$ will be discussed below).
This imposes the constraint on the
spin directions
within the
chains: ${\bf e}_{{\bf i} \in \ell}$$=$${\bf e}_{\ell}$.
The distribution function for
${\bf e}_{\ell}$$=$$(e^x_{\ell},e^y_{\ell},e^z_{\ell})$,
which is required in order to calculate the thermal (or orientational) averages,
can be taken in the form
$P({\bf e}_{\ell})$$\sim$$\exp (-h e^z_{\ell})$, corresponding to the
mean-field approximation for inter-chain interactions. Then, $\xi_{\bf ij}$$=$$1$
for ${\bf i}$ and ${\bf j}$ belonging to the same chain.
Our goal is to calculate the electronic
Green function $\widehat{\overline{\cal G}}$
(corresponding to the first two terms in $\widehat{\cal H}_{\bf ij}$),
averaged over $\{ {\bf e}_{\ell} \}$.
As the zeroth order approximation along this line,
all $\xi_{\bf ij}$ connecting neighboring zigzag chains
are replaced by their orientational averages,
$\xi_{\ell \ell \pm 1}$$\rightarrow$$\overline{\xi}$ \cite{deGennes}.
Corrections to this approximations can be constructed in a systematic way by
using the perturbation theory expansion with respect to fluctuations
$\Delta \xi$$=$$\xi_{\ell \ell \pm 1}$$-$$\overline{\xi}$ with
subsequent averaging over $\{ {\bf e}_{\ell} \}$. Then, $\widehat{\overline{\cal G}}$
has the following form,
after transformation to the reciprocal space ($k_y$) along $y$
and using matrix representation
for the orbital, chain, and inequivalent atomic indices:
$\widehat{\overline{\cal G}}_{k_y}(\varepsilon)$$=$$\left\{ 1 +
\sum_{n=2}^\infty \overline{(\Delta \xi)^n}
[ \widehat{\cal G}^0_{k_y}(\varepsilon)\widehat{t}^{\pm}_{k_y} ]^n \right\}\widehat{\cal G}^0_{k_y}(\varepsilon)$,
where $\widehat{\cal G}^0_{k_y}$ corresponds to the zeroth order DE Hamiltonian, and
$\widehat{t}^{\pm}_{k_y}$$\sim$$\delta_{\ell \ell \pm 1}$ is the super-matrix of transfer
interactions between NN chains.
In these (shorthanded) notations, $\Delta \xi$'s have the
same chain indices as $\widehat{t}^{\pm}_{k_y}$'s.
Since $\overline{\Delta \xi}$$=$$0$, the first-order
correction vanishes, and the matrices $\widehat{t}^{\pm}_{k_y}$,
which enters the matrix multiplication for higher $n$,
should have at least one common chain index in order to contribute to $\widehat{\overline{\cal G}}$.
Then, it is feasible to go up to $n$$=$$2$ and $3$.
In most cases, $n$$=$$2$ is already
sufficient \cite{comment.4}.
Finally, $\widehat{\overline{\cal G}}_{k_y}(\varepsilon)$ can be
transformed to the real space $\widehat{\overline{\cal G}}_{\bf ij}(\varepsilon)$.

  The free energy ${\cal F}$ is estimated using variational mean-field
approach~\cite{deGennes}. First, we calculate the DE
energy:
$$
E_{\rm DE} = -\frac{1}{\pi} {\rm Im} \sum_{\bf i} \int d \varepsilon
\ln \left[ 1 + \exp \left(\frac{\varepsilon-\varepsilon_F}{k_BT} \right) \right]
{\rm Tr}_L \widehat{\overline{\cal G}}_{\bf ii}(\varepsilon).
$$
with ${\bf i}$ running
over the unit cell shown in Fig.~\ref{fig.pattern}, $\varepsilon_F$ being the
chemical potential, and ${\rm Tr}_L$ denoting the trace over $L$.
Then, $E_{\rm DE}$ is combined with the SE energy and the entropy term, corresponding
to the molecular field $h$~\cite{deGennes}. This yields ${\cal F}(h)$, which
should be minimized with respect to $h$ or the AFM order parameter
$m_A$$=$$\tanh^{-1} h$$-$$h^{-1}$.

  The orbital polarization at the site $A$
(the difference of atomic populations of the $3x^2$-$r^2$ and $y^2$-$z^2$ orbitals, in
the oxygen frame) is given by
$$
{\cal O}_A=-\frac{1}{2\pi} {\rm Im} \int d \varepsilon
f_T(\varepsilon-\varepsilon_F) {\rm Tr}_L \{ (\sqrt{3}\widehat{\sigma}_x+\widehat{\sigma}_z)
\widehat{\overline{\cal G}}_{AA}(\varepsilon)\},
$$
where $f_T(\varepsilon-\varepsilon_F)$ is the Fermi
distribution function. (Similar parameter ${\cal O}_B$ related
with $\widehat{\Delta}_{B}$ at the site $B$ is of the order of $0.1$,
and only weakly depends on the
degrees of the JT distortion and the spin disorder). The
charge disproportionation
$$
{\cal C}=-\frac{1}{\pi} {\rm Im} \int d \varepsilon
f_T(\varepsilon-\varepsilon_F) {\rm Tr}_L \{
\widehat{\overline{\cal G}}_{AA}(\varepsilon)-
\widehat{\overline{\cal G}}_{BB}(\varepsilon) \}
$$
shows the degree of CO.
The stiffness of the FM zigzag chain is controlled by~\cite{Springer}
$$
J_{AB}^{\uparrow \uparrow}=-\frac{1}{2 \pi} {\rm Im} \int d \varepsilon
f_T(\varepsilon-\varepsilon_F) {\rm Tr}_L \{ \widehat{t}_{AB}
\widehat{\overline{\cal G}}_{BA}(\varepsilon) \}.
$$
In order to be stable against the thermal fluctuations $J_{AB}^{\uparrow \uparrow}$ should satisfy
the inequality:
$(J_{AB}^{\uparrow \uparrow}$$+$$J^S)$$\gg$$k_B T$.

\begin{figure}[h!]
\centering \noindent
\resizebox{7.0cm}{!}{\includegraphics{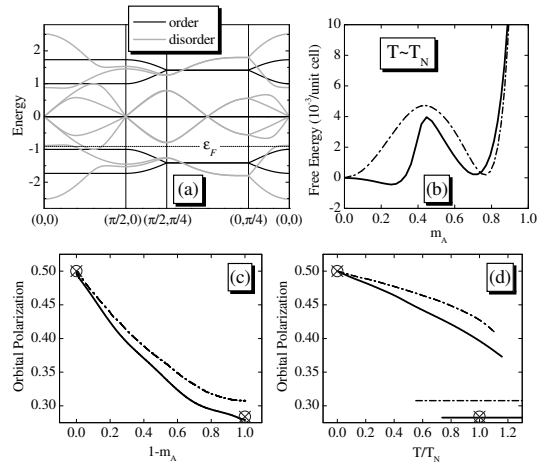}}
\caption{(a) Effects of inter-chain disorder on the electronic structure
for $\delta_{\rm JT}$$=$$0$. The disorder
closes the band gap and leads to the semi-metallic (SM)
behavior for small $m_A$.
(b) Free energy ${\cal F}$ against $m_A$ for $T$$\sim$$T_{\rm N}$
and $|J^S|$$=$$0.13$. The SM electronic structure
leads to the additional gain of the DE energy and stabilizes the new minimum
at $m_A$$=$$0$. (c) and (d) show the orbital polarization
${\cal O}_A$ against $m_A$ and $T$. The form of
${\cal F}(m_A)$ predetermines the first-order transition
and discontinuity of ${\cal O}_A$ (d). Dot-dashed lines, solid lines
and symbols $\otimes$ correspond to the zeroth, second and third order of
$\Delta\xi$, respectively. $T_{\rm N}$, obtained from
${\cal F}(0)$$=$${\cal F}(m_A)$, is $0.15$ and $0.10$ in the zeroth and second
order of $\Delta\xi$, respectively.}
\label{fig.JT00}
\end{figure}

  Results for $\delta_{\rm JT}$$=$$0$ are summarized
in Fig.~\ref{fig.JT00}.
The ordered AFM state is a band insulator~\cite{Hotta,PRL99,PRB01,Brink}.
Will it be insulating
in the case of the spin
disorder between the chains?
In the zeroth order of $\Delta\xi$, it is equivalent to the change of
$\xi_{\ell \ell \pm 1}$ in the DE Hamiltonian from $\overline{\xi}$$=$$0$ ($m_A$$=$$1$) to
$\overline{\xi}$$=$$\frac{2}{3}$ ($m_A$$=$$0$) \cite{deGennes}.
The corresponding electronic structure is \textit{semi-metallic} (SM), meaning that there
is a direct gap in each ${\bf k}$-point of the Brillouin zone.
However, a finite
overlap between different ${\bf k}$-points closes the real gap.
Thus, the DE mechanism alone is insufficient to make the disordered zigzag phase
a band insulator, though the situation is certainly close to it, and as we will see below
the insulating behavior can be easily produced by the JT distortion.

  The SM character of the electronic structure predetermines the type of the phase
transition at $T_{\rm N}$. The collapse of the band gap at small $m_A$ leads to the
gain of the DE energy and stabilizes an additional minimum of ${\cal F}$ at
$m_A$$=$$0$, which coexists with another minimum
corresponding to an insulating phase at finite $m_A$.
Thus, the phase transition will be of the first order,
which is
characterized by the abrupt change of ${\cal O}_A$.
${\cal C}$ vanishes without
distortion~\cite{PRB01}.
This scenario is qualitatively consistent with the behavior of NSMO
compounds. It also explains the observed phase coexistence around
$T_{\rm N}$~\cite{Mori,comment.1}, which may lead to the percolation and
the CMR effect~\cite{percolation}.
However, most of CO
manganites, like NCMO, exhibit a continuous second-order transition at $T_{\rm N}$.

\begin{figure}[h!]
\centering \noindent
\resizebox{7.0cm}{!}{\includegraphics{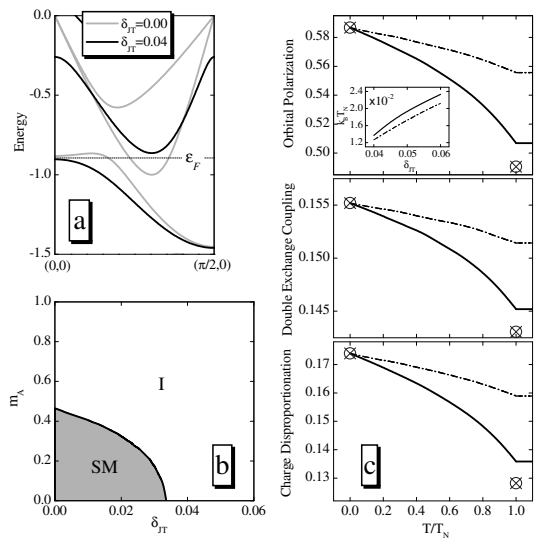}}
\caption{(a) The electronic structure of the disordered zigzag phase with and
without JT distortion. $\delta_{\rm JT}$
opens the band gap. (b) Phase diagram showing the regions of
semi-metallic (SM) and insulating (I) behavior versus $\delta_{\rm JT}$ and
$m_A$. (c) An example of temperature behavior
of ${\cal O}_A$, $J_{AB}^{\uparrow \uparrow}$, and ${\cal C}$
for $\delta_{\rm JT}$$\sim$$0.04$ and $|J^S|$$=$$0.09$. The
opening of the
band gap by $\delta_{\rm JT}$ leads to the second-order transition at
$T_{\rm N}$. The inset shows $T_{\rm N}$ versus $\delta_{\rm JT}$.
See Fig.~\protect\ref{fig.JT00} for other notations.}
\label{fig.JT04}
\end{figure}

  What is missing in the above picture? Below, we will argue that the difference
between NSMO and NCMO is controlled by the ratio $\lambda/\kappa$~\cite{Yunoki}, and
can be understood by considering the \textit{response} of the crystal structure to the
highly anisotropic magnetic structure. First, we estimate the force
$f$$=$$-$$\left. \partial E_{\rm DE}/\partial \delta_{\rm JT} \right|_{\delta_{\rm JT}=0}$
acting on the lattice degrees of freedom and coming from the electron-phonon term
in Eq.~(\ref{eqn:H})~\cite{comment.3}, which couples $\delta_{\rm JT}$ with the
parameters of orbital polarization. This yields (per one unit cell)
$f$$=$$-$$2 \lambda ({\cal O}_A$$+$${\cal O}_B)$, which is of the order of 10 eV.
Then, we note that even a modest distortion $\delta_{\rm JT}$$\sim$$0.04$~\cite{comment.5}
is sufficient to open the band gap in the disordered zigzag phase
(Fig.~\ref{fig.JT04}). Using $f$, we estimate the
spring constant $\kappa$ which would correspond to such equilibrium distortion:
$\kappa$$=$$-$$f/4 \delta_{\rm JT}$. This yields $\kappa$$\simeq$$65$ eV,
in agreement previous estimates~\cite{Millis}.
The temperature behavior corresponding to this ratio $\lambda/\kappa$
is
shown in Fig.~\ref{fig.JT04}c.
Since the system is insulating for all $m_A$, ${\cal F}(m_A)$ has only
one minimum for each $T$. Therefore, the transition at $T_{\rm N}$
will be of the second order. Finite $\delta_{\rm JT}$
enhanced the orbital polarization
${\cal O}_A$ and gives rise to the small charge disproportionation ${\cal C}$.
Both parameters are affected by the spin disorder and monotonously
decrease with $T$, that explains similar intensity change of the CO/OO
superlattice peaks in the neutron scattering experiment \cite{Sternlieb}.
$T_{\rm N}$, obtained from
$\left. \partial^2 {\cal F}/\partial m_A^2 \right|_{m_A=0}$$=$$0$,
depends on the JT distortion.
For $\delta_{\rm JT}$$=$$0.04$ and $|J^S|$$=$$0.09$,
it is of the order of $120$ K, which is close to the experimental
value $110$ K observed in LSMO~\cite{Tokura,Sternlieb}.
Then, the ration is $(J_{AB}^{\uparrow \uparrow}$$+$$J^S)/k_B T_{\rm N}$ is
about 4, meaning that the FM coupling within the chains remains robust
even around $T_{\rm N}$. This justifies the proposed model
of inter-chain disorder.

  In summary, CO
in manganites is a magnetic phenomenon.
The transition from the AFM ground state to the paramagnetic state
goes through the intermediate phase, in which
the FM order within the zigzag chains coexists with the inter-chain disorder.
This implies that the multi-critical behavior in CMR manganites~\cite{Tomioka}
is the result of competition between different \textit{magnetic} phases,
which is only masked by spurious CO.
Details of the phase transition at $T_{\rm CO}$
present an interesting and so far the least understood
part of the problem. We would like to stress two points.
(i) The transition should be discontinuous, so that the change of the
electronic structure should be sufficient to open the gap and self-stabilize
the anisotropic magnetic state.
(ii) The transition should be accompanied by the lattice distortion,
which is a necessary precondition for insulating behavior above $T_{\rm N}$.
Both features are clearly
seen in the experiment~\cite{Tokura}.
An interesting question, which still
remains to be understood
is how the FM zigzag chains will break. An important step in this direction
would be the explanation of
\textit{incommensurate} CO superstructures
observed just
below $T_{\rm CO}$~\cite{Chen}.

  I thank all member of Tokura group for the great deal of inspiration
by numerous experimental stories about CO in manganites.


\begin{thebibliography}{10}
\bibitem{comment.2}
The Berry phases do not contribute to the
one-dimensional spin disorder.
\bibitem{comment.4}
Note that
$\overline{(\Delta \xi)^2}$$=$$\frac{1}{18}$ and
$\overline{(\Delta \xi)^3}$$=$$-$$\frac{1}{135}$
when $h$$\rightarrow$$0$.
All $\overline{(\Delta \xi)^n}$ vanish in the oppsite limit
$h$$\rightarrow$$\infty$.
\bibitem{comment.1}
The first order transition is a necessary precondition for the
phase coexistence. The effect itself is driven by the
configurational mixing entropy.
\bibitem{comment.3}
The first corrections to $t$ and $J^S$ upon the change of $\delta_{\rm JT}$
are proportional to $\delta^2_{\rm JT}$, provided that
$d_L$$+$$d_S$$=$$const$. Both are small in comparison
with $\kappa \delta^2_{\rm JT}$.
\end{thebibliography}
\end{document}